\documentclass[sigconf, screen]{acmart}
\setcopyright{none}
\renewcommand\footnotetextcopyrightpermission[1]{}

\usepackage{booktabs}
\usepackage{array}
\usepackage{colortbl}
\usepackage{multirow}
\usepackage{xspace}

\newcommand{\xx}{\ensuremath{\times}\xspace}
\newcommand{\sch}[1]{\mbox{[#1]}}
\newcommand{\pifive}{\ensuremath{\pi_{0.5}}\xspace}
\newcommand{\mmsq}{mm\textsuperscript{2}\xspace}

\definecolor{insight}{HTML}{4F7D5C}
\definecolor{insightT}{HTML}{ECF4EE}
\newcounter{insight}
\newcommand{\insight}[1]{\par\addvspace{6pt}\noindent\refstepcounter{insight}{\color{insight}\vrule width 2pt}%
  {\fboxsep0pt\colorbox{insightT}{\vbox{\kern2.5pt\hbox{\kern5pt\parbox[b]{\dimexpr\linewidth-11pt\relax}{%
  \textbf{Insight~\theinsight.}~#1}\kern4pt}\kern2pt}}}\par\prevdepth\dp\strutbox\addvspace{6pt}}

\usepackage{tikz}
\DeclareRobustCommand{\cn}[1]{\tikz[baseline=(n.base)]\node[circle, fill=black!85, text=white, inner sep=0pt,
  minimum size=1.15em, font=\bfseries\footnotesize] (n) {#1};}

\makeatletter
\newcommand{\resdef}[2]{\@namedef{res@#1}{#2}}
\newcommand{\res}[1]{\@ifundefined{res@#1}{\textcolor{red}{??}\@latex@warning{Result `#1' undefined}}{\@nameuse{res@#1}}}
\makeatother
\resdef{ms.2334}{8.02}
\resdef{hz.2334}{125}
\resdef{array.2334}{5.20}
\resdef{over.2334}{20}
\resdef{linearpct.2334}{94}
\resdef{attnpct.2334}{22}
\resdef{array.w8}{13.31}
\resdef{over.w8}{14}
\resdef{pct.dramrate}{88}
\resdef{ms.n15}{5.16}
\resdef{ms.pi25}{2.70}
\resdef{mj.2334}{78}
\resdef{mjchip.2334}{35}
\resdef{dramshare.2334}{51}
\resdef{wfifty.2334}{3.9}
\resdef{mjchip.w8}{91}
\resdef{pct.e2334w8}{31}
\resdef{pct.e24w8}{18}
\resdef{pj.dram}{5.5}
\resdef{x.drammul}{79}
\resdef{pj.dramdie}{3.4}
\resdef{pj.dramphy}{2.1}
\resdef{w.standby}{0.43}
\resdef{ms.thorbf16}{31.5}
\resdef{ms.thor2334}{13.3}
\resdef{x.thor2334}{1.7}
\resdef{mj.thor2334}{543}
\resdef{pct.thorbytes}{25}
\resdef{pct.thortime}{9}
\resdef{ms.thorfixed}{3.2}
\resdef{pct.thorfixed}{65}
\resdef{pct.thorroof2}{48}
\resdef{pct.thorint82}{21}
\resdef{mw.n16}{830}
\resdef{gbbf16.n16}{6.6}
\resdef{ctx.n16}{121--207}
\resdef{gmacstep.n16}{43}
\resdef{mw.n15}{415}
\resdef{gbbf16.n15}{3.3}
\resdef{ctx.n15}{566}
\resdef{gmacstep.n15}{21}
\resdef{mw.pi05}{311}
\resdef{gbbf16.pi05}{6.2}
\resdef{ctx.pi05}{660}
\resdef{gmacstep.pi05}{18}
\resdef{xattn.lo}{seven}
\resdef{xattn.hi}{nine}
\resdef{gaincalls.lo}{45}
\resdef{gaincalls.hi}{60}
\resdef{pct.gainfirst}{23}
\resdef{pct.spreadsana}{54}
\resdef{alloc.lo}{1.1}
\resdef{alloc.hi}{2.3}
\resdef{hard.stream}{five}
\resdef{hard.attn}{three}
\resdef{hard.lo}{17}
\resdef{hard.hi}{30}
\resdef{mm2.phy}{14.4}
\resdef{mm2.placed}{4.6}
\resdef{die.placed}{19}
\resdef{pct.diephy}{76}
\resdef{pct.period2334}{39}
\resdef{s2.free}{12.1}
\resdef{s2.periods}{four}
\resdef{s2.hz}{12.5}
\resdef{thor.policymj}{2944}
\resdef{x.s2thor}{1.2}
\resdef{x.headthor}{3.9}
\resdef{s2.rcpolicymj}{405}
\resdef{x16ch.w8}{1.3}
\resdef{x16ch.2334}{3.1}
\resdef{ns.odrpath}{0.88}
\resdef{pct.odrtime}{0.4}
\resdef{gmac.2334}{169}
\resdef{pct.odrbank}{51}
\resdef{pct.odrdie}{5.0}
\resdef{pct.odrdiepim}{19.4}
\resdef{x.odruncoded}{1.4}
\resdef{pct.odrrtl}{2.4}
\resdef{pct.odrtraffic}{3.6}
\resdef{sota.ours.bits}{7.98}
\resdef{x.flashrt}{1.7}
\resdef{xj.flashrt}{9.1}
\resdef{x.ditpaw8}{1.3}
\resdef{hz.ditpan15}{60}
\resdef{x.ditpan15}{1.7}
\resdef{hz.ditpapi25}{84}
\resdef{x.ditpapi25}{4.4}
\resdef{x.pim}{8.7}
\resdef{bpw.n16.2334}{7.98}
\resdef{xbpw.n16.2334}{8.0}
\resdef{bpw.n16.24}{4.7}
\resdef{pct.nv23}{6}
\resdef{bpw.nv23scales}{0.5}
\resdef{pct.entgap}{2.3}
\resdef{pct.codemin}{55}
\resdef{x.decpeak}{1.8}
\resdef{bpc.decpeak}{2000}
\resdef{gb.master}{0.85}
\resdef{gb.copies}{3.3}
\resdef{kept.2334}{five}
\resdef{kept.25}{five}
\resdef{task.mean}{$-$1}
\resdef{task.max}{2}
\resdef{xbpw.task.2334}{7.8}
\resdef{jerk.lo}{6}
\resdef{jerk.hi}{16}
\resdef{jump.lo}{7}
\resdef{jump.hi}{16}
\resdef{xjerk.g1}{1.9}
\resdef{h.h100}{1700}

\begin{document}

\title{Hardware Acceleration of Vision-Language-Action Models with Weight Precision Rising over the Flow Steps}

\author{Robin Geens, Simon Haepers, Jonas Crols, Chao Fang, Marian Verhelst}
\affiliation[obeypunctuation=true]{\institution{\rule{0pt}{14pt}MICAS, KU Leuven}, \country{Belgium}}

\begin{abstract}
Vision-language-action (VLA) models such as GR00T control robots through an action head: a diffusion transformer that turns noise into a chunk of actions in a few flow-matching steps. Every step re-reads all of the head's weights from DRAM but applies each weight to only about fifty tokens. Inference is therefore memory-bound, and its latency and energy are largely determined by the total number of transferred weight bits. We show that later flow steps attenuate the quantization errors of earlier steps, so the weight precision of an action head should rise over the steps. With plain per-channel rounding and without calibration or training, widths rising over the steps from 2 to 4 bits preserve the closed-loop success of robot policies. These widths transfer 25\% fewer bits than 4-bit weights at every step. However, we show that on Jetson AGX Thor, current GPU kernels cannot exploit this per-step precision due to fixed overheads. We therefore present a programmable accelerator whose latency and energy both scale with the transferred weight bits, because it keeps its DRAM link saturated throughout an inference at every width. It decodes entropy-coded weights for bit-serial sub-arrays at the DRAM rate. Because the width changes per step, a run-time allocator divides these sub-arrays between the linear layers and the attention to balance their workloads. The same sub-arrays are thus capable of running action heads of different families and sizes, as well as a vision-language model. On the DRAM side, a requantizer and an entropy encoder at each bank let one stored 8-bit copy replace the copies of all widths. Simulated with LPDDR5X timing, our RTL implementation runs the GR00T N1.6 action head in \res{ms.2334}\,ms per inference with 2, 3, 3 and 4 bits in its four flow steps. With a copy per width and 50\% of Thor's memory bandwidth, the accelerator runs this schedule \res{x.flashrt}\xx faster than Thor in NVFP4, at \res{xj.flashrt}\xx less energy and equal closed-loop success.

\end{abstract}

\maketitle
\pagestyle{plain}

\section{Introduction}
\label{sec:intro}

\begin{figure}[t]
\centering
\includegraphics[width=\columnwidth]{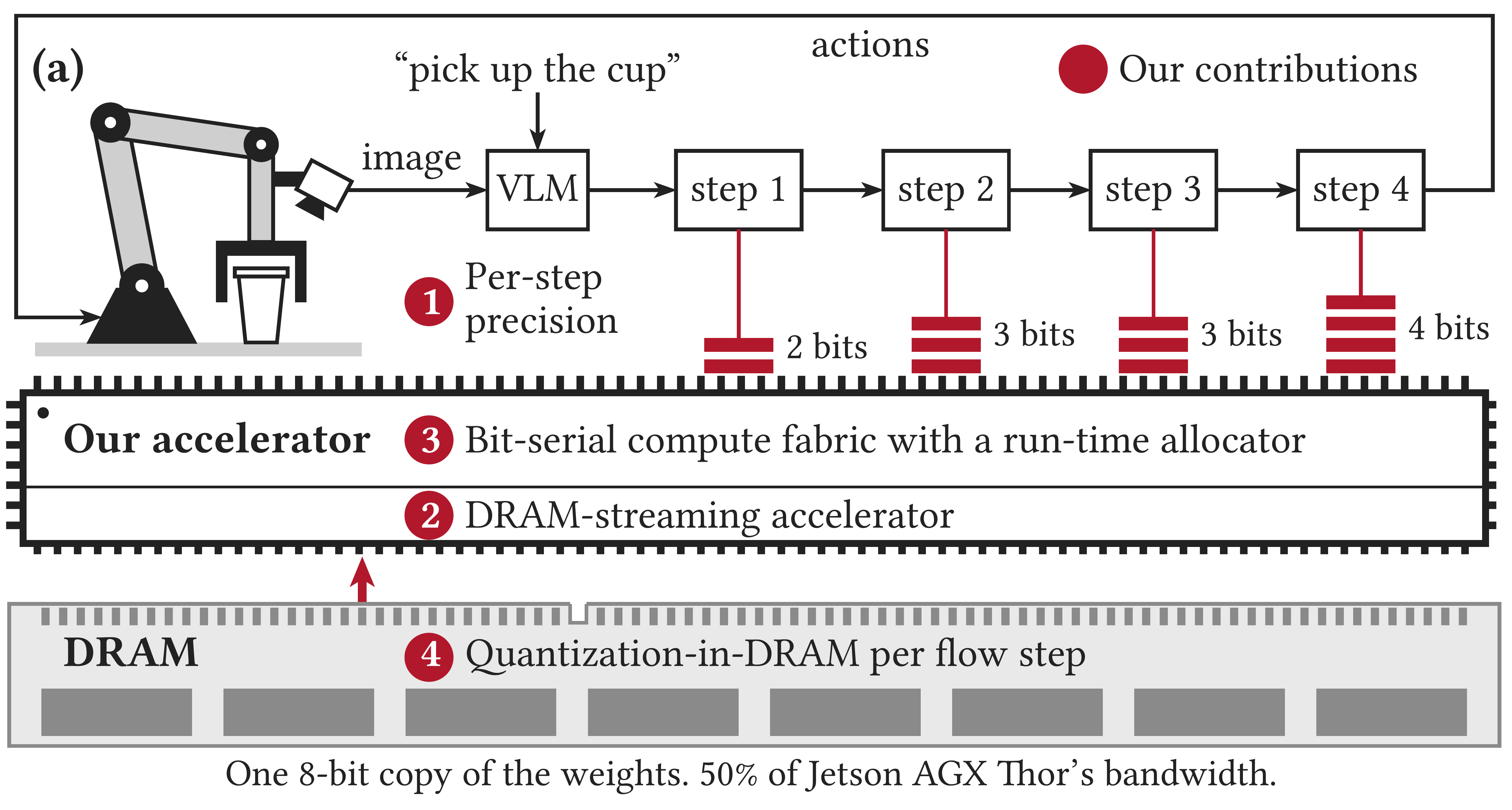}\\[3pt]
\includegraphics[width=\columnwidth]{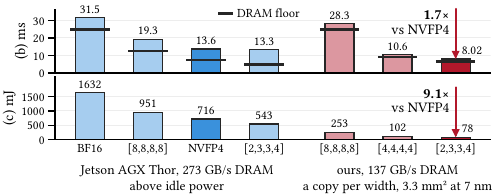}
\caption{(a) A robot's policy with per-step weight precision in System~1 on our accelerator. (b) Time and (c) energy per inference of the GR00T N1.6 action head,
without the VLM and with a copy of the weights per width.}
\label{fig:overview}
\end{figure}

Vision-language-action (VLA) models are the state-of-the-art in general-purpose robot manipulation~\cite{openvla,pi0,pi05,gr00tn1}. Most VLAs split the policy in two, as Figure~\ref{fig:flow} shows: a vision-language model (VLM), \emph{System~2}, interprets the scene and instruction, encoding them into tokens. Conditioned on these tokens and the robot state, a diffusion-transformer (DiT) action head, \emph{System~1}, turns noise into a chunk of actions in a limited number of flow-matching steps~\cite{flowmatching,rectifiedflow,dit}.

System~1 is the latency-critical part of a dual-system policy, since it closes the control loop once per chunk. Even when the robot executes one chunk while the next is computed~\cite{rtc,smolvla}, the latency of System~1 bounds how often the robot can react to a new observation. To react before every action, System~1 must run at the control rate, such as 50\,Hz~\cite{pi0}. System~2, in contrast, tolerates more latency and runs at 10\,Hz or less~\cite{gr00tn1,helix}. Like recent accelerators for robot policies~\cite{ditpa,deltoris}, this work therefore accelerates System~1.

The action head is memory-bound: every flow step re-reads all of its weights but applies each weight to only the roughly fifty tokens of one chunk. As a result, the GR00T N1.6 head streams gigabytes of weights per inference in BF16, as Table~\ref{tab:models} shows. Today, robot policies run on edge GPUs~\cite{helix,vlahwbench,isaacgr00t} such as NVIDIA's Jetson AGX Thor, which draws up to 130\,W~\cite{jetsonthor}. NVIDIA reports 10.9\,Hz for GR00T N1.6 on Thor with TensorRT, and the action head takes over half of each inference~\cite{isaacgr00t}. One inference per action at 50\,Hz would need more than the GPU's whole memory bandwidth in BF16. Moreover, the GPU of an autonomous mobile robot can draw over twice the power of its motors~\cite{pnav}. A lower energy per inference thus extends how long the robot runs on one charge.

Weight quantization reduces the bits per weight. Prior work quantizes current policies to \emph{uniform} 4-bit weights, with the same width at every flow step~\cite{foldquantvla,loopq}. Below 4 bits, however, uniformly rounded weights lose closed-loop success with and without calibration.

We observe that the flow steps do not need equal precision. Because an action head refines a wide range of starting noise into almost the same action, its later steps also attenuate the quantization error of an early step. An error of the last step, in contrast, reaches the robot in full. The weight width should thus rise over the flow steps. Plain per-channel rounding of the released model then suffices, because the coarsest widths fall on the steps that tolerate them. For example, the \emph{schedule} \sch{2,3,3,4} gives two bits to the first step and four to the last. Without calibration data or training, it preserves the released model's success for three VLAs. Schedules with falling widths, in contrast, lower the success rate on all three.

A GPU, however, cannot exploit the reduced DRAM traffic below 4 bits. Figure~\ref{fig:overview}b shows that fixed overheads dominate the time per inference that we measured on Jetson AGX Thor (Section~\ref{sec:gpu}). First, the attention and the normalization run between the linear layers while no weights stream from DRAM. Second, the linear layers process narrow weights more slowly than the DRAM delivers them. Exploiting per-step precision thus requires hardware that runs the attention and the normalization in parallel with the weight stream and keeps up with the DRAM at every width. Moreover, the hardware must adapt to its dynamic workload, because the width changes per step and the attention's share is model- and context-dependent. Our contributions, marked in Figure~\ref{fig:overview}, are:
\begingroup
\settowidth{\leftmargini}{\cn{1}}\addtolength{\leftmargini}{\labelsep}
\begin{itemize}
\item[\cn{1}] An analysis showing that later flow steps attenuate early quantization errors and that plain rounding to
  rising widths therefore preserves closed-loop success with fewer bits than uniform widths (Section~\ref{sec:time}).
\item[\cn{2}] A DRAM-streaming accelerator whose latency and energy scale with the transferred weight bits, because it
  decodes entropy-coded weights while keeping its DRAM channels saturated at every width
  (Sections~\ref{sec:arch-overview}--\ref{sec:arch-code}).
\item[\cn{3}] A bit-serial compute fabric with a run-time allocator that divides it between the linear layers and the
  attention, so that it runs different action heads and a VLM
  (Sections~\ref{sec:arch-array}--\ref{sec:arch-vlm}).
\item[\cn{4}] A Quantization-in-DRAM system that requantizes and encodes the weights at every bank, so that a deployment
  needs only one stored 8-bit copy to switch between schedules (Section~\ref{sec:ondram}).
\end{itemize}
\endgroup

\noindent Section~\ref{sec:eval} evaluates the schedules in closed-loop simulations of robot tasks. It also evaluates the RTL of the accelerator, whose outputs are bit-exact with the action head of these simulations. The energy comes from gate-level simulations that replay the value change dump (VCD) of an RTL inference on the synthesized netlists with an estimated clock tree.

\section{Background and Motivation}
\label{sec:background}

\begin{figure}[t]
\centering
\includegraphics[width=\columnwidth]{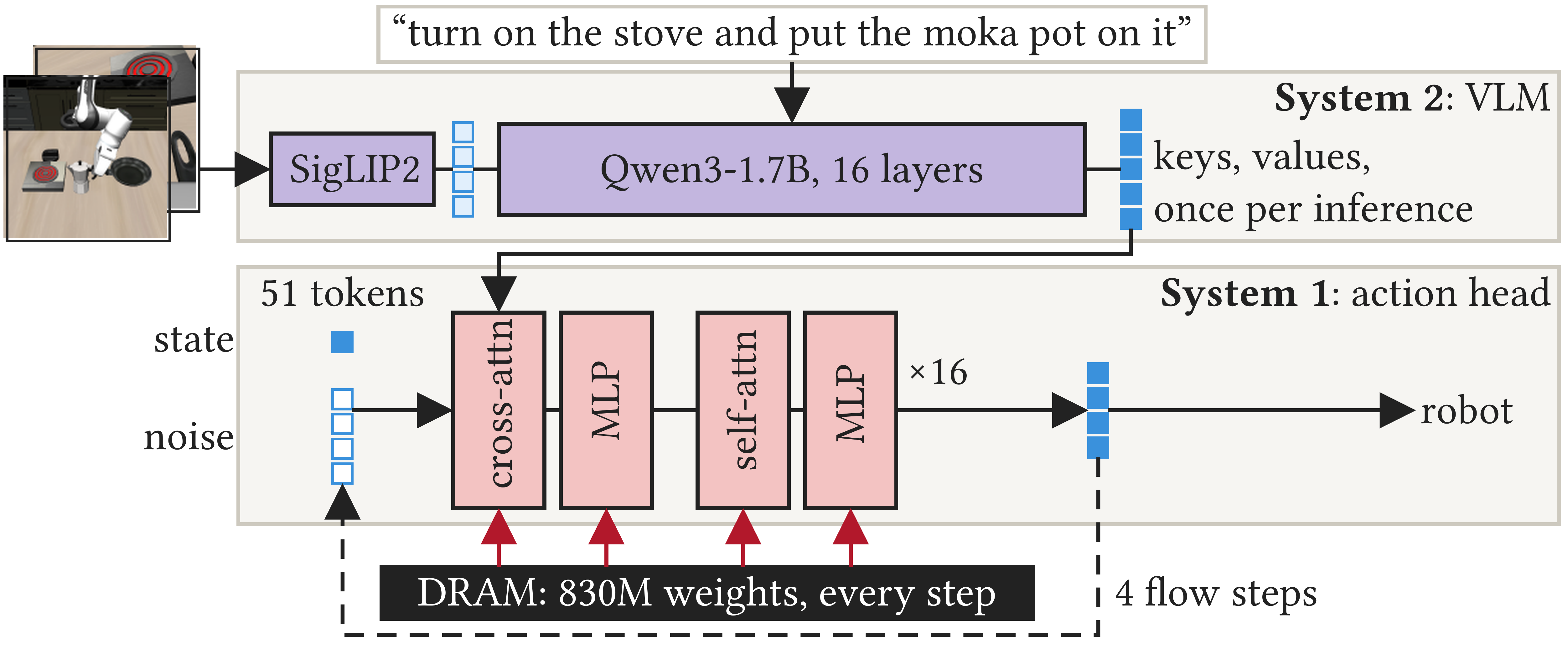}
\caption{GR00T N1.6 on a LIBERO-Long task.}
\label{fig:flow}
\end{figure}

This section describes VLA action heads such as System~1 in Figure~\ref{fig:flow} and shows why their inference is memory-bound.

\subsection{Flow-Matching Action Heads}

A flow-matching action head learns a velocity field $\mathbf{v}_\theta(\mathbf{x}, t \mid \mathbf{c})$ that carries Gaussian noise $\mathbf{x}_0$ at $t = 0$ to an action chunk at $t = 1$. The field is conditioned on a context $\mathbf{c}$ of System~2's features and the robot's state~\cite{flowmatching,pi0}. Inference integrates the field with $N$ Euler steps of size $\Delta t = 1/N$:
\begin{equation}
\mathbf{x}_k = \mathbf{x}_{k-1} + \Delta t \, \mathbf{v}_\theta(\mathbf{x}_{k-1}, t_{k-1} \mid \mathbf{c}), \qquad k = 1, \dots, N.
\label{eq:euler}
\end{equation}
Throughout this paper, an \emph{inference} is one such integration. In GR00T, the velocity network is a diffusion transformer~\cite{dit} whose blocks alternate cross-attention to System~2's features with self-attention over the chunk. Each block receives the time $t_{k-1}$ through adaptive LayerNorm (AdaLN) and ends in a two-layer MLP~\cite{gr00tn1,gr00tn16}. Since the keys and values of System~2's features are constant across steps, System~2 hands them over once per inference. Table~\ref{tab:models} lists the action heads of three state-of-the-art open-weight VLAs.

\begin{table}[t]
\caption{The action heads that the accelerator runs.}
\label{tab:models}
\footnotesize
\setlength{\tabcolsep}{2pt}
\begin{tabular*}{\columnwidth}{@{\extracolsep{\fill}}lrrrrrrr@{}}
\toprule
 & & \multicolumn{2}{c}{Tokens} & \multicolumn{2}{c}{Per flow step} & \multicolumn{2}{c}{Per BF16 inference} \\
\cmidrule(lr){3-4}\cmidrule(lr){5-6}\cmidrule(l){7-8}
 & Blocks & Chunk & Context & Weights [M] & MACs [G] & Steps & Read [GB] \\
\midrule
GR00T N1.6 & 32 & 51 & \res{ctx.n16} & \res{mw.n16} & \res{gmacstep.n16} & 4 & \res{gbbf16.n16} \\
GR00T N1.5 & 16 & 49 & \res{ctx.n15} & \res{mw.n15} & \res{gmacstep.n15} & 4 & \res{gbbf16.n15} \\
\pifive & 18 & 50 & \res{ctx.pi05} & \res{mw.pi05} & \res{gmacstep.pi05} & 10 & \res{gbbf16.pi05} \\
\bottomrule
\end{tabular*}
\end{table}

\subsection{Memory-Bound Inference}
\label{sec:bound}

Since a robot serves one observation at a time, every flow step applies all weights of the action head to the roughly fifty tokens of a single chunk. With about fifty multiply-accumulate operations (MACs) per weight and step, the operational intensity lies far below the roofline ridge point of GPU tensor cores and accelerator arrays. The inference is therefore memory-bound and takes at least the \emph{DRAM floor}, the time to read its bytes once at the peak DRAM bandwidth. Its energy is at least that of these reads, which can cost \res{x.drammul}\xx more energy per bit than an INT8 multiplication in 7\,nm~\cite{micronlp5x,samsungpnm,tpuv4i}.

We measure the weight traffic of an inference in \emph{bit-steps}, the bits read per weight summed over all $N$ steps. A schedule $[b_1, \dots, b_N]$ of per-step widths reads $\sum_k b_k$ bit-steps, against 64 for BF16 at four steps. For example, the schedule \sch{2,3,3,4} reads 12 bit-steps, 5.3\xx fewer than BF16. Entropy-coded weights (Section~\ref{sec:arch-code}) take even fewer bit-steps than the sum of the widths. 
\insight{Reading weight bits from DRAM dominates the latency and energy of an inference, so reducing its bit-steps can reduce both proportionally.}

\section{Per-Step Sensitivity of Action Heads}
\label{sec:time}

\subsection{Quantization Method}
\label{sec:quantizer}

\begin{figure}[t]
\centering
\includegraphics[width=\columnwidth]{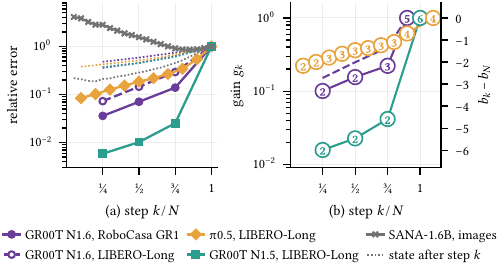}
\caption{(a) Errors of the output and of the state after step $k$ when only step $k$ of $N$ is quantized to 3 bits, relative to quantizing only the last step.
(b) Gain $g_k$, the ratio of the two errors, with each step's width at the minimum of Equation~\ref{eq:rule}.}
\label{fig:gains}
\end{figure}

For each width of a schedule, we quantize the released BF16 model by per-channel round-to-nearest, without calibration data, training or tuning for a specific robot or task. Each output channel has one scale, and all blocks of a step share one width. At 3 bits and more, we scale each row so that its absolute maximum maps to $2^{b-1}-1$ and round every weight to a two's-complement integer. At 2 bits, a mid-rise code without a zero level uses the levels $\pm d/2$ and $\pm 3d/2$, with $d$ fitted to the row by least squares.

The schedule covers all linear layers of the DiT blocks. The small layers that embed the noisy actions as input tokens and that decode the output tokens into a velocity stay 8-bit. The AdaLN modulation depends only on the step index and remains a constant vector per step. The inputs of the linear layers are rounded to INT8 with one scale per token, which maps the token's absolute maximum to 127. Their width stays constant over the steps because the weights carry nearly all of the DRAM traffic at batch size one. For simplicity, we also keep the activations of the attention, its queries, keys and values, at 8 bits. The softmax, LayerNorm, GELU and the residual stream stay in BF16 or FP32.

\subsection{Attenuation of Early Quantization Errors}
\label{sec:gains}

Every flow step reads the same weights, but the steps do not necessarily need the same precision. An action head's output depends on the observation but barely on the starting noise. Across noise draws, the actions of the three policies differ by only a few percent of the noise magnitude. The flow thus maps a wide range of starting points to almost the same action. It must therefore attenuate any deviation of the intermediate states, including the quantization error of an early step.

To test this intuition, we quantize the weights of only one step at a time to 3 bits and compare the final action with that of the unquantized model. We run this comparison open loop on \res{gaincalls.lo} to \res{gaincalls.hi} model calls per benchmark in Figure~\ref{fig:gains}, recorded in closed-loop episodes. The error is the root-mean-square difference between the two actions over four noise draws, averaged geometrically over the calls. Every action dimension enters this error normalized by its range in the training data, so that no single dimension dominates it. Figure~\ref{fig:gains}a shows that the error due to quantizing the last step exceeds that of the first by one to two orders of magnitude. A uniform width thus spends most of its bits on steps whose errors barely reach the action.

Two effects can make the last step more sensitive than the first. Quantizing step $k$ perturbs its velocity and leaves an error $s_k$ in the state $\mathbf{x}_k$ (Equation~\ref{eq:euler}). The later steps carry this state error to the action, where it becomes the output error $e_k$. First, quantizing a late step may leave a larger state error. Second, the later steps may attenuate the state error of an early step before it reaches the action. To separate the two effects, we define the \emph{gain} of step $k$ as $g_k = e_k / s_k$, with both errors taken from the same run. The state after the last step is the output action, so $g_N = 1$. On all three policies, the state error resulting from quantizing the last step is only two to three times that of the first step, as Figure~\ref{fig:gains}a shows. Figure~\ref{fig:gains}b shows that the remaining steps, in contrast, attenuate an error of the first step to \res{pct.gainfirst}\% or less. This attenuation thus accounts for most of the difference in sensitivity between the first and the last step. The larger state error of the last step accounts for the rest.

Action heads differ in this respect from image diffusion models, whose output depends strongly on the starting noise. The images of the flow-matching DiT SANA-1.6B~\cite{sana}, for example, differ across noise draws by \res{pct.spreadsana}\% of the noise magnitude. Figure~\ref{fig:gains}a shows that the output error of SANA-1.6B exceeds its state error at every step but the last. Its remaining steps thus amplify an early error. Per-step quantizers for image diffusion accordingly give the noisiest steps the most weight bits, as in TMPDiff~\cite{tmpdiff} and TASQ~\cite{tasq}. An action head, in contrast, needs its bits on the last steps.

\insight{Because an action head's output barely depends on the starting noise, its early steps tolerate more quantization than its late ones, and bit widths should rise over the steps.}

\subsection{Precision Schedules from the Gains}
\label{sec:rule}

The gains also indicate how to distribute a given number of bit-steps over the steps. If the errors of the steps propagate linearly to the action, are uncorrelated and halve with every additional bit, the squared action error of a schedule approximately satisfies
\begin{equation}
E(b_1,\dots,b_N) \propto \sum_{k=1}^{N} \left(g_k \, 2^{-b_k}\right)^2.
\label{eq:rule}
\end{equation}
For $B = \sum_k b_k$ bit-steps, Lagrange multipliers give the minimum of $E$ at $b_k = \log_2 g_k + (B - \sum_j \log_2 g_j)/N$. This yields a rule of thumb: a step receives one more bit per doubling of its gain. Figure~\ref{fig:gains}b gives the resulting width difference to the last step, $b_k - b_N$, on its right axis. With a gain of about 0.1, the first step of GR00T N1.6 receives about three bits fewer than the last. The total $B$ determines the traffic and is a choice of the deployment. Section~\ref{sec:eval-acc} applies the rule at the same $B$ as uniform 3-bit weights, 12 bit-steps for both GR00T policies and 30 for \pifive. The rule's widths preserve or nearly preserve success, whereas uniform 3-bit weights lose success on GR00T N1.6 and \pifive.

\insight{An action head's gains determine its step widths as a rule of thumb, so finding a schedule needs no closed-loop search.}

\section{Motivation for Dedicated Hardware}
\label{sec:gpu}

\begin{figure}[t]
\centering
\includegraphics[width=\columnwidth]{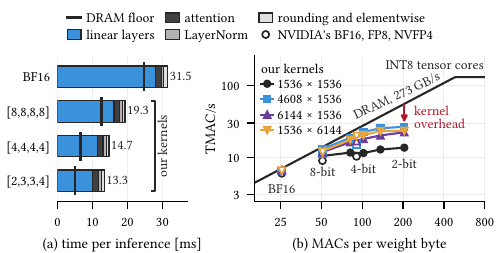}
\caption{GR00T N1.6 on Jetson AGX Thor: (a) time per inference by kernel type and (b) the roofline of its linear layers.}
\label{fig:thor}
\end{figure}

The schedules of Section~\ref{sec:time} reduce the weight bytes that an inference reads from DRAM. On an edge GPU such as Thor, the time per inference falls in proportion to these bytes only if the DRAM remains the bottleneck at every width. To run the schedules on Thor, we write CUDA kernels for linear layers with 2- to 8-bit weights. Our kernels store a copy of every layer per width. We compare the action head on our kernels with the released BF16 action head, whose linear layers run through NVIDIA's cuBLAS library.

Figure~\ref{fig:thor}a shows that on our kernels, \sch{2,3,3,4} takes \res{ms.thor2334}\,ms per GR00T N1.6 inference, against \res{ms.thorbf16}\,ms for the released BF16 action head. From \sch{4,4,4,4} to \sch{2,3,3,4}, however, the weight bytes fall by \res{pct.thorbytes}\% but the time only by \res{pct.thortime}\%. The kernels themselves are not the limitation: Figure~\ref{fig:thor}b shows that at equal bytes, they are about as fast as NVIDIA's FP8 and NVFP4 matrix multiplications~\cite{nvfp4} or faster.

Two factors keep the time far above the DRAM floor. First, the attention, the LayerNorm and the rounding of the activations take about \res{ms.thorfixed}\,ms per inference at every width, \res{pct.thorfixed}\% of the DRAM floor of \sch{2,3,3,4}. Second, the roofline in Figure~\ref{fig:thor}b shows that the linear layers fall further below the DRAM roof as their weights narrow. At 2 bits, even the largest layers reach at most \res{pct.thorroof2}\% of this roof and \res{pct.thorint82}\% of the peak compute of the INT8 tensor cores. Since neither DRAM nor tensor cores limit these layers, the gap to the DRAM roof is kernel overhead that does not decrease as the weights narrow.
\insight{Below 4 bits, current GPU kernel stacks are bound by fixed overheads, so sub-4-bit precision needs hardware whose entire latency scales with the bit-steps.}

To remove the first factor, dedicated hardware must apply the normalization, rounding, activation functions and residual additions as the activations enter or leave the linear layers. It must also run the attention in parallel with the weight stream. To remove the second factor, it must process the weights faster than the DRAM delivers them, at every width from 2 to 8 bits. Moreover, it must adapt to a dynamic workload. The operational intensity of the linear layers changes with the width of each step, and the share of the attention changes between models and contexts. Section~\ref{sec:arch} presents such an accelerator.

\section{Accelerator Architecture}
\label{sec:arch}

\begin{figure*}[t]
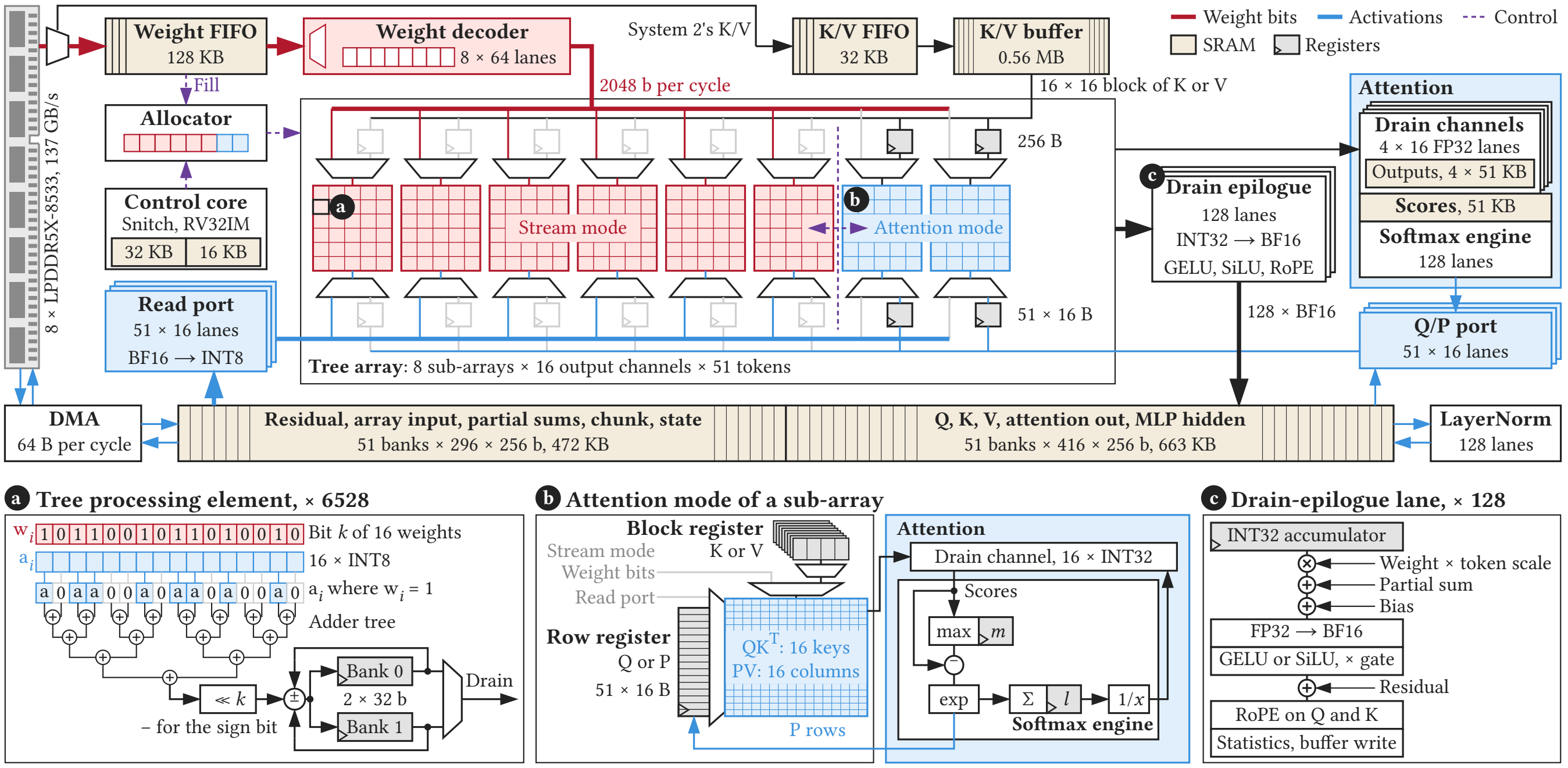

\centering
\includegraphics[width=\textwidth]{figures/arch_system.pdf}\\[4pt]
\includegraphics[width=\textwidth]{figures/arch_units.pdf}
\caption{The accelerator with sub-arrays in stream and attention mode. Each sub-array's inputs of its unused mode are grayed.}
\label{fig:arch}
\end{figure*}

Figure~\ref{fig:arch} shows the accelerator. Every flow step streams all weights of the action head from DRAM through a FIFO and a decoder into the array. Only the activations of one chunk, the keys and values of the current layer and the constants reside on chip. We design the accelerator to keep the DRAM channels saturated at every width, so that the time of an inference approaches its DRAM floor.

Because the best schedule differs between policies and benchmark suites, one datapath must run each flow step at any width from 2 to 8 bits. An output-stationary bit-serial array~\cite{stripes,loom} meets this requirement. Consuming one bit of each weight per cycle, it runs a step of any width as the same loop of $b$ cycles. Since streaming a step's weights also takes time proportional to $b$, the array keeps the DRAM channels saturated at every width.

\subsection{Memory Interface}
\label{sec:arch-overview}

The proposed accelerator has a 128-bit memory interface of eight x16 LPDDR5X-8533 channels with 8\,GB of capacity and a peak bandwidth of 136.5\,GB/s, 50\% of Thor's. We target 50 action-head inferences per second, one per action at a 50\,Hz control rate~\cite{pi0}. This target suffices for most applications, since policies typically execute several actions of each chunk~\cite{pi0,gr00tn1,rtc}. The weight FIFO lets the DRAM continue to stream weights while the accelerator processes layers that require no weights, such as LayerNorm and the attention. A DMA engine moves the inputs of an inference and the keys and values that a layer produces. System~2's keys and values arrive through the K/V FIFO.

\subsection{Entropy-Coded Weight Streams}
\label{sec:arch-code}

\begin{figure}[t]
\centering
\includegraphics[width=\columnwidth]{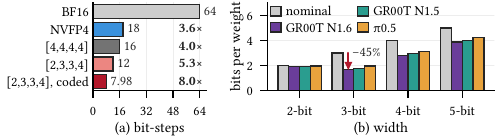}
\caption{(a) Bit-steps of a GR00T N1.6 inference. (b) Bits read per weight at each width with the entropy code.}
\label{fig:coding}
\end{figure}

Every linear layer streams from DRAM at the width of its step, as one contiguous stream located through an address table per layer and width. The DRAM either stores one such stream per width or produces it from one stored 8-bit copy, as proposed in Section~\ref{sec:ondram}. Per-channel rounding scales each row by its largest weight, which lies far out in the tail of the distribution. Most weights are thus small compared with the step between two levels, and 56\% of the 3-bit weights of GR00T N1.6 round to zero. We therefore Huffman-code the streams, with code words of at most 12 bits. Each layer and width has one canonical code, stored at the start of its stream. Since every code word takes at least one bit, a code word covers a pair of neighboring weights at 2 and 3 bits. A decoder between the weight FIFO and the array decodes the code words in parallel, faster than the DRAM delivers them at every width. Because the code is lossless, coding does not alter the closed-loop success of the quantized policies.

Figure~\ref{fig:coding}b shows the coded bits per weight, which come within \res{pct.entgap}\% of the rounded weights' entropy for every policy and width. The entropy code thus further reduces the bit-steps of a \sch{2,3,3,4} inference of GR00T N1.6 from 12 to \res{bpw.n16.2334}, as Figure~\ref{fig:coding}a shows.

Microscaling formats such as NVFP4~\cite{nvfp4} lower the rounding error at 4 bits with an 8-bit scale per block of 16 weights instead of one per row. Each block maps its largest weight to the top level, so its weights spread over all levels and barely compress. Weights rounded per channel to 5 bits, in contrast, concentrate near zero, so the entropy code saves more than one bit per weight. Coded per-channel rounding at 5 bits therefore reads fewer bits than coded NVFP4 with its scales, at a lower output error.

\insight{Microscaling formats lower the output error but are less compressible, so entropy-coded weights reach the same error in fewer bits with one scale per row.}

\subsection{Flexible Bit-Serial Fabric}
\label{sec:arch-array}

The compute fabric is a bit-serial tree array of processing elements (PEs), 128 output channels by 51 tokens, enough for the chunk of every action head in Table~\ref{tab:models}. 
Per cycle, the array consumes bit $k$ of 16 weights of every output channel. As Figure~\ref{fig:coding}b shows, the code stores a weight in as little as \res{pct.codemin}\% of its width. At the 1\,GHz clock, the decoder thus emits up to \res{x.decpeak}\xx the 1092 bits that the DRAM delivers per cycle, about \res{bpc.decpeak} bits. We therefore size the array to consume 2048 weight bits per cycle, so that it keeps up with the decoder at every width. Since every weight bit meets 51 tokens, the array performs 104448 bit-MACs per cycle, each of which multiplies one weight bit with an INT8 activation. With per-step precision, the compute that the weight stream requires changes between flow steps, because the decoder emits the most bits per cycle at the most compressible widths. The compute of the attention varies between model families. For example, \pifive performs \res{xattn.lo} to \res{xattn.hi} times as many attention MACs per flow step as GR00T N1.6. To maximize flexibility and hardware utilization, the weight stream and the attention share the array. We split the array into eight sub-arrays of 16 output channels each, which a run-time allocator divides between the two.

Figure~\ref{fig:arch}a shows a PE, which serves one output channel and one token. Per cycle, it multiplies bit $k$ of 16 weights of its output channel with the INT8 activations of its token at the same 16 input channels. An adder tree sums the 16 products. The PE shifts the sum to the significance of bit $k$ and adds it into a 32-bit accumulator, or subtracts it for the two's-complement sign bit. Because one accumulator serves 16 input channels, the PE takes less area and energy per bit-MAC than a bit-serial PE that accumulates every product separately. The array computes a layer in \emph{passes}, each over one group of output channels. Two accumulator banks let one pass drain while the next fills. Since the allocator assigns each sub-array to the weight stream or to the attention, a sub-array runs every pass in one of two modes, \emph{stream mode} or \emph{attention mode}.

\textbf{Stream mode.} For the projections, a sub-array takes the decoded weights of its 16 output channels and the activations from the read port, as Figure~\ref{fig:arch} shows. The read port feeds the activations of all 51 tokens from the token buffers to every sub-array of the stream, 16 input channels per token and cycle. It rounds the activations of each token to INT8 with a scale from the token's largest magnitude. The drain epilogue (Figure~\ref{fig:arch}c) applies the elementwise operations that a GPU runs as separate kernels. It processes one token row per cycle while the array computes the next pass, so these operations do not stall the weight stream. A LayerNorm unit of the same width normalizes the token rows with statistics that the drain gathers, so that normalization occupies no sub-array. The output projection is pipelined with the attention in four tiles of input channels, so that earlier tiles run in parallel with the attention of later heads.

\textbf{Attention mode.} For the attention, the keys or values take the place of the weights, and the queries or the softmax probabilities P take that of the activations. A sub-array takes one bit of each key or value per cycle from a $16 \times 16$ block in its block register, as Figure~\ref{fig:arch}b shows. Its row register holds the matching 16 elements of the query or of P for each of the 51 tokens. A pass of $QK^\top$ puts 16 keys on the output channels, and a pass of $PV$ puts 16 columns of the head dimension there. A Q/P port rounds Q to 8 bits and loads Q or the two bytes of P into the row registers. Each tile of a head's keys passes in order through $QK^\top$, an online-softmax engine and $PV$. Because the integer accumulators hold both products exactly, the attention matches a dedicated attention unit bit for bit.

\textbf{Control core.} To run action heads of different sizes and architectures, a Snitch RISC-V core~\cite{snitch} runs each inference as a program that issues commands to the engines without delaying them. Through double-buffered command registers, the core writes the next command of an engine while the engine still executes the current one. Each command thus starts in the cycle that its engine becomes free. A command can even start before the command that produces its inputs has finished. Each of its reads then waits at a read gate only until the pass that produces the read's data has drained. A dependent GPU kernel, by contrast, reads its inputs only after the whole kernel before it has finished. The attention thus runs in parallel with the weight stream instead of adding to the fixed overheads of Section~\ref{sec:gpu}. A new schedule, step count or context length changes only the commands that the program issues. A new model family is thus a new program, so a single fabricated chip can run GR00T N1.6, GR00T N1.5 and \pifive.

\subsection{Run-Time Sub-Array Allocation}
\label{sec:arch-alloc}

\begin{figure}[t]
\centering
\includegraphics[width=\columnwidth]{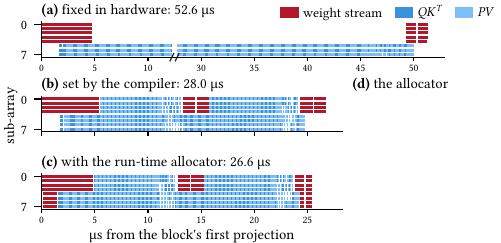}\llap{\includegraphics[scale=0.1]{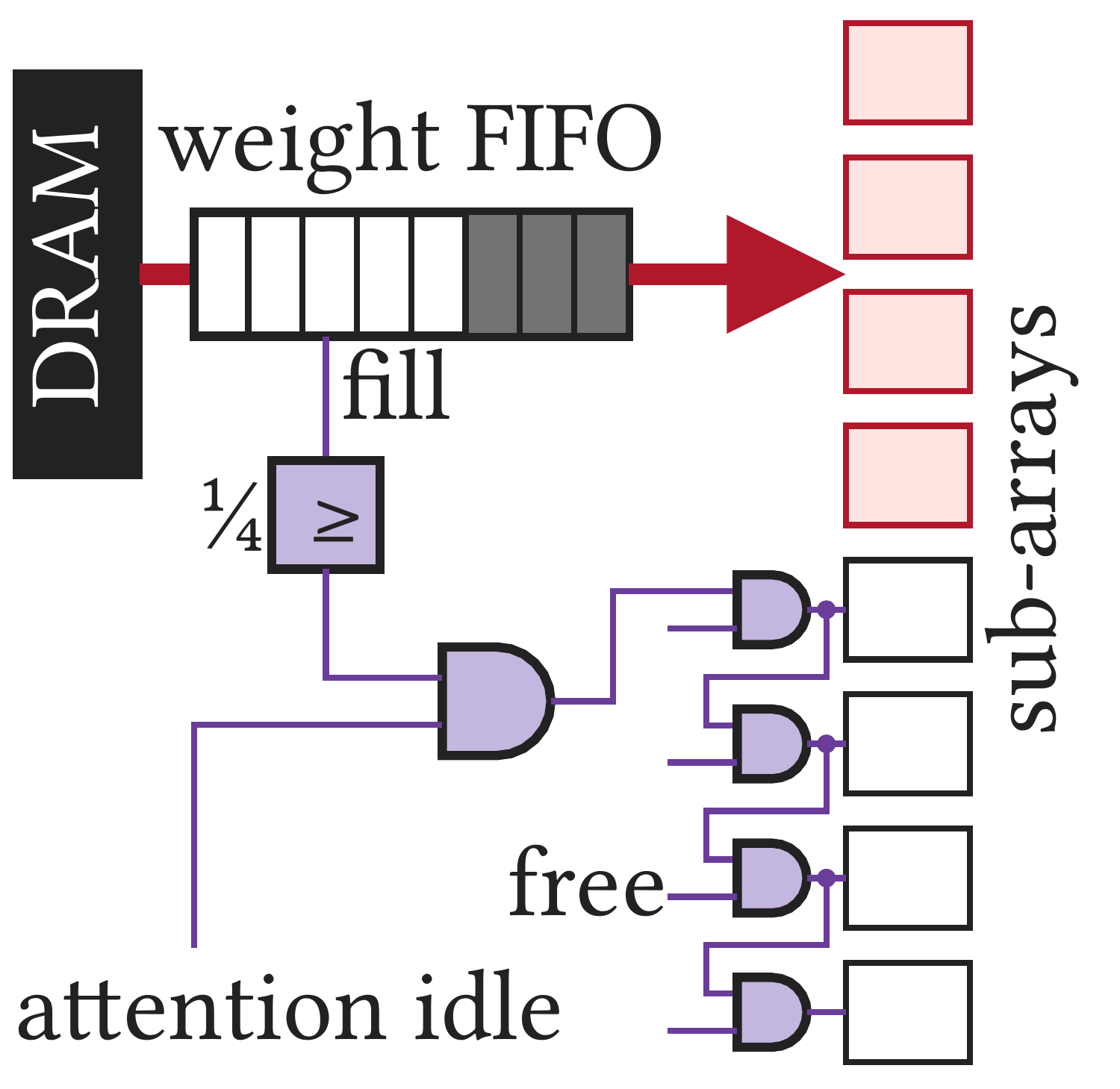}}
\caption{The eight sub-arrays in a \pifive block at the 2-bit step.}
\label{fig:allocation}
\end{figure}

To keep the DRAM channels saturated, the attention runs in parallel with the weight stream, on the sub-arrays that the projections do not need. At each width, a projection needs its \emph{sub-array minimum}, the fewest sub-arrays that keep up with the DRAM. A hardwired partition, as in Figure~\ref{fig:allocation}a, would be the same for every policy and width. Since the needs of the weight stream and the attention vary with both, it would leave the sub-arrays of one idle while the other lacks sub-arrays. Alternatively, the compiler can allocate the sub-arrays statically, as in Figure~\ref{fig:allocation}b. Each projection then takes as many sub-arrays as its sub-array minimum, starting from the first, and the attention takes the others. Whenever the projections halt, the DRAM keeps filling the weight FIFO and idles once it is full. For example, a tile of the output projection multiplies the outputs of a group of attention heads, so it halts until their attention ends. After the halt, the FIFO regains room only if the projections consume the weights faster than the DRAM delivers them. At their sub-array minimum, however, they consume them about as fast. With the static allocation, the FIFO thus remains full after the first halt, and the DRAM idles during every later halt.

Our run-time allocator (Figure~\ref{fig:allocation}d) instead decides at the start of every pass how many sub-arrays the weight stream takes, while each attention pass takes the free ones. A weight FIFO at least 25\% full shows that the DRAM runs ahead of the projections. If the attention then has no pass ready and no softmax in flight, the stream's pass also takes the free sub-arrays beyond its minimum, up to the first busy one. Otherwise, it runs on its minimum.

Figure~\ref{fig:allocation} compares the three options on a \pifive block. The hardwired partition with the smallest worst-case slowdown over the three action heads gives \res{hard.stream} sub-arrays to the weight stream and \res{hard.attn} to the attention. Even with this partition, the inferences of the three action heads take \res{hard.lo} to \res{hard.hi}\% longer than with the static allocation. The stream's \res{hard.stream} sub-arrays are too few at the 3-bit steps of GR00T N1.6, and the attention's \res{hard.attn} are too few on the other two heads. On \pifive, the stream's sub-arrays idle for most of the block, as Figure~\ref{fig:allocation}a shows. With the run-time allocator, the stream takes all eight sub-arrays before the attention starts and the free ones during the last passes of the output projection. The block thus ends earlier than with the static allocation, as Figures~\ref{fig:allocation}b and~c show. Over whole inferences, the accelerator runs the three action heads \res{alloc.lo} to \res{alloc.hi}\% faster with the allocator than with the static allocation.
\insight{At run time, the fill of the weight FIFO shows which fraction of the shared array should be allocated to the weight stream.}

\subsection{Running System~2}
\label{sec:arch-vlm}

The same accelerator can also run the VLM of GR00T N1.6 bit-exactly. This System~2 consists of SigLIP2 on each camera's image and the first 16 layers of Qwen3-1.7B. Its 1.33 billion weights apply to hundreds of tokens per layer instead of the 51 tokens of the array. Run in groups of 51 tokens, System~2 would re-stream every weight for every group. It therefore runs weight-stationary: each weight tile is loaded once into a buffer per sub-array and replayed for every group. The activations stream in from DRAM, whereas the partial sums stay on chip. Since every layer writes its outputs to DRAM, System~2 can pause at any layer boundary, so that action-head inferences can run in between.

\subsection{Quantization-in-DRAM}
\label{sec:ondram}

\begin{figure}[t]
\centering
\includegraphics[width=\columnwidth]{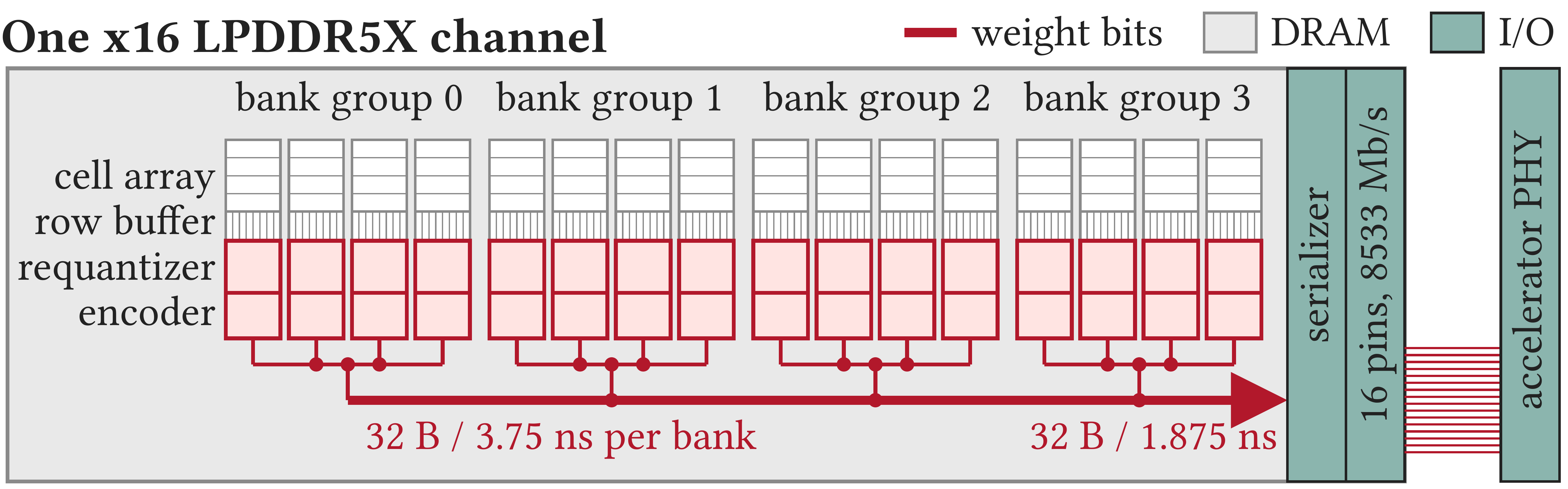}
\caption{Quantization-in-DRAM in one LPDDR5X channel.}
\label{fig:drambank}
\end{figure}

A deployment that switches between step counts and precision schedules may use any width from 2 to 8 bits. Our per-channel rounding uses a separate scale for every width, so a narrower weight is not a truncation of a wider one. Such a deployment therefore stores a copy of every width in DRAM, \res{gb.copies}\,GB for the copies of GR00T N1.6 against \res{gb.master}\,GB for one 8-bit copy. The $b$-bit weight, however, depends only on the 8-bit weight and the $b$-bit scale of its row. Every width can thus be derived from one stored 8-bit copy while the weights are read. The capacity saved can hold the weights of a larger model or of System~2.

\insight{Since per-channel rounding needs only a weight and its row's scale, one stored 8-bit copy can generate every width without extra DRAM capacity.}

We therefore propose Quantization-in-DRAM: a DRAM system that stores one 8-bit copy and requantizes and encodes the weights at every bank, as illustrated in Figure~\ref{fig:drambank}. At every step, the requantizer behind each row buffer rounds the stored bytes to the step's width, and the encoder Huffman-codes the result.

\textbf{Requantizer and encoder.} At the 2-bit step, the banks read at least 4\xx as many bits as leave the DRAM. The banks of a channel, however, share one bus to its interface, with the same bandwidth as the interface. A requantizer at the interface would thus use at most a quarter of its bandwidth. Every bank therefore holds its own requantizer and encoder, so that only coded weights cross the shared bus.

\textbf{Multiplication in the DRAM.} Quantization-in-DRAM adds only a requantizer and an encoder to each bank. Processing-in-memory DRAMs instead add MAC units to the banks, as Samsung's HBM-PIM~\cite{hbmpim} and LPDDR5X-PIM~\cite{samsungpim} do, so that no weight crosses the memory interface. Logic in a DRAM process, however, is about 10\xx less dense than in a logic process~\cite{upmem}, so MAC units at every bank are large. They also perform one MAC per weight read, whereas GR00T N1.6 applies every weight read to 51 tokens, so they would take far longer than the accelerator. Quantization-in-DRAM therefore leaves the MACs to the accelerator.

\section{Experimental Methodology}
\label{sec:method}

\begin{figure*}[t]
\centering
\includegraphics[width=\textwidth]{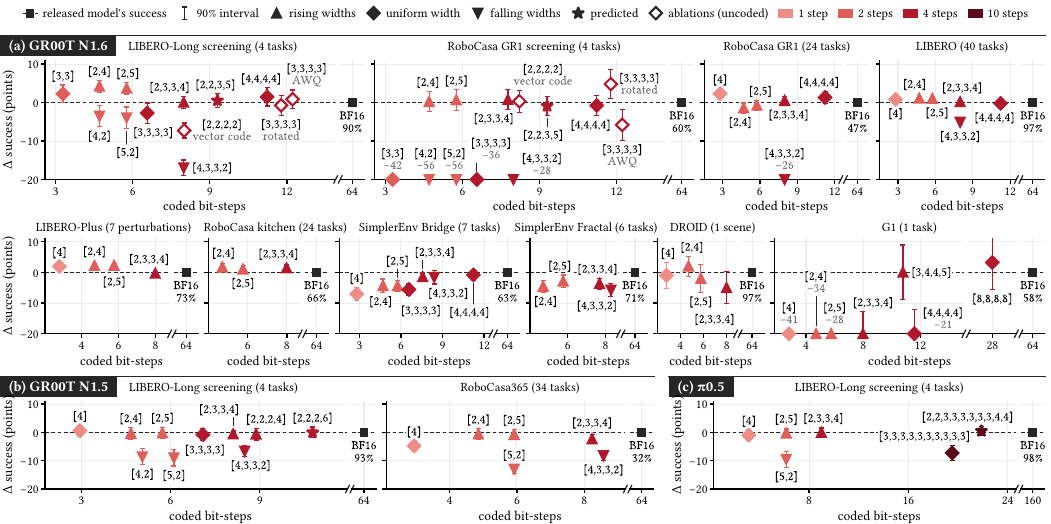}
\caption{Change in closed-loop success relative to the released model over the entropy-coded bit-steps.}
\label{fig:accuracy}
\end{figure*}

We evaluated the closed-loop success of the precision schedules in simulated environments (roughly \res{h.h100} H100 hours) and the time, area and energy of the accelerator from its RTL. In closed loop, the simulated robot executes a task over many model calls.

\textbf{Policies.} GR00T N1.6-3B~\cite{gr00tn16} runs its base checkpoint on RoboCasa GR1 and the RoboCasa kitchen. A third-party LIBERO fine-tune~\cite{gr00tn16libero} runs LIBERO and LIBERO-Plus. NVIDIA's Bridge, Fractal and G1 checkpoints run on their benchmarks, and its DROID checkpoint on one scene of the DROID simulation. GR00T N1.5~\cite{gr00tn15} runs NVIDIA's LIBERO-Long checkpoint on LIBERO-Long and the RoboCasa team's checkpoint on RoboCasa365. \pifive~\cite{pi05} runs LeRobot's LIBERO fine-tune~\cite{lerobot} at its ten steps. Every quantized policy runs in a released inference harness, simulator and control loop~\cite{isaacgr00t,lerobot}. Only its action head is replaced by one with the data types and roundings of the accelerator.

\textbf{Benchmarks.} To compare many schedules, we composed two screening benchmarks of four tasks each. The first consists of the four LIBERO-Long~\cite{libero} tasks on which the released model succeeded least, since such tasks reveal a loss with few episodes. The second consists of four RoboCasa GR1 tabletop tasks~\cite{robocasa,gr00tn1}, two from each of its two task families. We also evaluated eight full suites in which we ran all tasks of RoboCasa GR1, LIBERO, LIBERO-Plus~\cite{liberoplus}, the RoboCasa kitchen, SimplerEnv~\cite{simplerenv,bridgev2,rt1} Bridge and Fractal, and G1. Of RoboCasa365, we ran the 34 target tasks on which its checkpoint was trained. A task runs at least 50 episodes per seed. We report the change in success rate against the released model as the mean over the tasks, each weighted equally. The 90\% interval of this change extends 1.645 standard errors to either side. Where both policies start each episode from the same state, the variance of a task's change comes from the per-episode differences. Otherwise, it comes from the binomial variances of both success rates. A schedule \emph{preserves success} when this interval lies above $-3$ points, following the equivalence margin of Sujit et~al.~\cite{onestep}.

\textbf{RTL simulation.} The accelerator was written in Chisel~\cite{chisel}, parameterized in the sizes of its array, token buffers, FIFOs and ports, and simulated in Verilator~\cite{verilator}. The cycle-accurate DRAM simulator Ramulator~2~\cite{ramulator2} models its LPDDR5X-8533 channels under JEDEC timing~\cite{lpddr5x}. Each simulated inference replays a model call recorded in closed loop. Its outputs are bit-exact with an executable model that follows the roundings of the closed-loop simulations.

\textbf{Area.} Every block of the accelerator was synthesized with Cadence Genus in TSMC 7\,nm at 1\,GHz. The SRAM came from the foundry's memory compiler, and each block's clock tree was estimated in Cadence Joules. The accelerator area excludes long wires, the pads and the physical interfaces (PHYs), whose area we estimated from a die shot~\cite{m2dieshot}.

\textbf{Energy.} We simulated the same netlists and clock trees at gate level with FlashReplay in Cadence Joules. The simulation replays the VCDs of a LIBERO inference at \sch{2,3,3,4} and of System~2 from the RTL, at the typical corner of 0.75\,V and 25\,$^\circ$C. We added the reads, writes and leakage of the SRAM macros from the foundry's memory-compiler tables. From Micron's LPDDR5X datasheet~\cite{micronlp5x} and Samsung's breakdown of LPDDR5 energy~\cite{samsungpnm}, a bit read from the DRAM costs \res{pj.dram}\,pJ. Of these, \res{pj.dramdie}\,pJ go to the DRAM's row activation, read path and output drivers, and \res{pj.dramphy}\,pJ to the PHY and memory controller of the accelerator. We added the DRAM's standby and refresh power, \res{w.standby}\,W for the eight channels at the datasheet's maximum currents, over the RTL time of each inference.

\textbf{Baselines.} On Thor, we measured three implementations of the GR00T N1.6 action head as replayed CUDA graphs: the released BF16 head, FlashRT's NVFP4 head~\cite{flashrt} and our kernels. Every implementation runs after a warm-up, with the GPU and memory clocks pinned at their maximum. We measured the energy at the input of the Thor module, which also feeds its CPU, memory and regulators, and subtracted the module's idle power. For System~2, we also measured NVIDIA's VLM backbone, as the median of 20 synchronized calls. The open-source cycle model of DiTPA~\cite{ditpa} runs the GR00T N1.6 action head with its skips disabled. For a fair comparison, every action-head time and energy excludes the projection of System~2's keys and values, which we count in System~2.

\section{Evaluation}
\label{sec:eval}

This section evaluates the rising precision \cn{1} in closed loop, the accelerator's latency, energy and area \cn{2}, its fabric on other action heads and the VLM \cn{3}, and Quantization-in-DRAM \cn{4}.

\subsection{Closed-Loop Task Success}
\label{sec:eval-acc}

\textbf{Rising bit order.} In Figure~\ref{fig:accuracy}, the GR00T N1.6 schedules of 12 nominal bit-steps show that the order of the bits decides whether a schedule preserves success. Only the rising \sch{2,3,3,4} preserves success on both screening benchmarks. Uniform 3-bit weights and the falling \sch{4,3,3,2} lose success on both. On GR00T N1.5 and on \pifive at four steps instead of ten, \sch{2,3,3,4} also preserves success, whereas every falling schedule loses success.

\textbf{Rule of thumb.} From the measured gains, the rule of thumb of Section~\ref{sec:rule} predicts \sch{2,2,3,5} for GR00T N1.6, \sch{2,2,2,6} for GR00T N1.5 and \sch{2,2,3,3,3,3,3,3,4,4} for \pifive. The schedules for GR00T N1.5 and \pifive preserve success, and that for GR00T N1.6 narrowly misses the margin on RoboCasa GR1. The gains measured on a policy thus predict a schedule that preserves or nearly preserves success, without a search over schedules in closed loop.

\textbf{Full benchmark suites.} Figures~\ref{fig:accuracy}a and~b give every full suite we evaluated. \sch{2,3,3,4} preserves success on \res{kept.2334} of the eight suites. The rising order thus holds beyond the screening benchmarks.

\textbf{Step count.} At two steps, the rising \sch{2,4} preserves success on both screening benchmarks. The uniform \sch{3,3} and the falling \sch{4,2} of equal nominal bit-steps, in contrast, lose success on RoboCasa GR1. The order of the bits thus matters at two steps as well as at four. Whether fewer steps preserve success, however, depends on the suite. \sch{2,5} and \sch{2,4} each preserve success on \res{kept.25} of the eight suites. The step count is therefore a program input.

\subsection{Trajectory Quality}
\label{sec:eval-traj}

\textbf{Smoothness of the commands.} A success rate alone could hide noise that the rounded weights add to the commands, since a simulated controller absorbs jitter that a real arm would execute. On real robots, abrupt changes of velocity between action chunks also cause jerky motion and shift the observed dynamics away from the training data~\cite{act,rtc}. We therefore measured the jerk of the commanded arm positions, their third time derivative, and the jump of their velocity between consecutive chunks. On RoboCasa GR1, LIBERO and the RoboCasa kitchen, \sch{2,3,3,4} lowers the jerk by \res{jerk.lo} to \res{jerk.hi}\% and the jump by \res{jump.lo} to \res{jump.hi}\% relative to the released model. Every 90\% interval of these changes lies below zero. On G1, where \sch{2,3,3,4} loses success, its jerk is \res{xjerk.g1}\xx that of the released model. On these four suites, the success rate thus hides no loss of smoothness.

\textbf{Episode length.} A quantized policy that preserves its success rate may still need more inferences per episode. On the suites where \sch{2,3,3,4} preserves success, its successful episodes change in length by \res{task.mean}\% on average and by at most \res{task.max}\% relative to the released model. Per task, \sch{2,3,3,4} thus still reads at least \res{xbpw.task.2334}\xx fewer weight bits than the released model, against \res{xbpw.n16.2334}\xx fewer per inference.

\insight{Quantization can lengthen a policy's episodes, so its time and energy must also be compared per task.}

\subsection{Calibration-Free Quantization}
\label{sec:eval-quant}

We compared the plain per-channel rounding of Section~\ref{sec:quantizer}, which needs no calibration data or training, with more elaborate quantization methods from prior work. None of these methods preserves success at fewer bit-steps than plain rounding to a rising precision.

\textbf{Nested code.} Plain rounding of every width preserves more success than the nested code of TASQ~\cite{tasq}. The nested code reads the $b$ most significant bits of one stored 8-bit copy instead of rounding to $b$ bits with a separate scale. Without TASQ's training, the nested code read as \sch{2,3,3,4} succeeds in almost none of the RoboCasa GR1 episodes. The rounded widths, in contrast, preserve the success of the released model. Nested codes thus work only when trained for the nesting, as in TASQ and Matryoshka Quantization~\cite{matquant}.

\textbf{Calibration.} Neither rotation~\cite{quarot} nor AWQ scaling~\cite{awq} lets uniform 3-bit weights preserve success on both screening benchmarks. Figure~\ref{fig:accuracy}a shows that rotation preserves it only on RoboCasa GR1 and AWQ only on LIBERO-Long. The rising \sch{2,3,3,4} reads the same nominal bit-steps with plain rounding and preserves success on both. GPTQ~\cite{gptq} updates the columns not yet rounded so that a layer's output on calibration data changes least. It still succeeds less often than plain rounding of the same schedule.

\textbf{Microscaling.} NVFP4~\cite{nvfp4} stores an 8-bit scale per block of 16 weights instead of one per row. With this scale and integer levels at each step's width, \sch{2,3} at two steps preserves success on LIBERO-Long, as \sch{2,4} does with plain rounding. Uncoded, both read 6 bit-steps, since the block scales add half a bit per weight at each step. Coded, however, \sch{2,4} reads \res{bpw.n16.24} bit-steps, \res{pct.nv23}\% fewer than the block-scaled \sch{2,3}, whose scales alone take \res{bpw.nv23scales} bit-steps. A coded weight stream thus needs only one scale per row, whereas block scales would add a multiplication of every adder-tree sum by its block's scale.

\textbf{Vector code.} A uniform 2-bit vector code, as in Deltoris~\cite{deltoris}, replaces each group of four input weights by an 8-bit codebook index. Each layer has one codebook, built from the weights alone. The code reads as many bit-steps as coded \sch{2,3,3,4}, but Figure~\ref{fig:accuracy}a shows that it loses success on LIBERO-Long. Its codebook also expands every weight back to 8 bits. The code would thus take the array time of uniform 8-bit weights, \res{array.w8}\,ms against \res{array.2334}\,ms for \sch{2,3,3,4}. Rounded weights, in contrast, keep the arithmetic as narrow as the bits read.

\subsection{Latency}
\label{sec:eval-time}

\begin{figure}[t]
\centering
\includegraphics[width=\columnwidth]{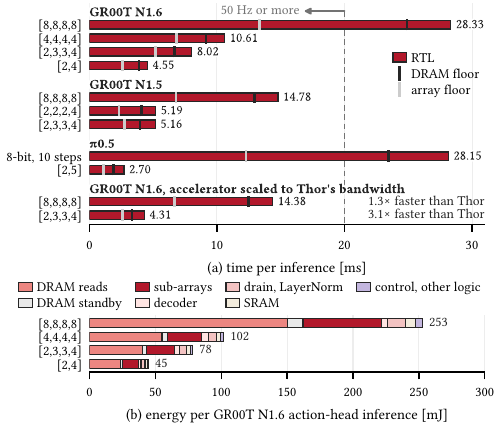}
\caption{Evaluation of the accelerator on DiT action heads.}
\label{fig:results}
\end{figure}

On a GPU, the time per inference moves further above the DRAM floor as the weights narrow (Section~\ref{sec:gpu}). Figure~\ref{fig:results}a compares the RTL time of every inference on the accelerator with its DRAM and array floors, the times at the DRAM and compute roofs of a roofline. The DRAM floor (defined in Section~\ref{sec:bound}) includes the keys and values of System~2. The \emph{array floor} is the time that the array takes for the weight stream at 2048 weight bits per cycle. On GR00T N1.6, the time stays \res{over.w8} to \res{over.2334}\% above the DRAM floor at every width, and \sch{2,3,3,4} takes \res{ms.2334}\,ms per inference, or \res{hz.2334}\,Hz. The DRAM floor ignores refreshes and row switches, which cut the bandwidth to \res{pct.dramrate}\% of the peak.

Unlike on a GPU, the attention, the LayerNorm and the rounding of the activations add little time on the accelerator. In a \sch{2,3,3,4} inference of GR00T N1.6, the RTL traces show the attention operations in flight for \res{attnpct.2334}\% of the time, all of it in parallel with the weight stream. Linear layers are in flight for \res{linearpct.2334}\% of the inference, and only LayerNorm and DMA commands take the rest. The linear layers also stay DRAM-bound as their weights narrow, because the array floor lies below the DRAM floor for every action head and schedule. The time per GR00T N1.6 inference thus scales with the bit-steps, down to \sch{2,4}, the narrowest schedule that we evaluated.

\subsection{Energy and Area}
\label{sec:eval-energy}

Figure~\ref{fig:results}b breaks down the energy per inference into the DRAM reads, the DRAM's standby and the parts of the chip. \sch{2,3,3,4} costs \res{mj.2334}\,mJ per inference, or \res{wfifty.2334}\,W at 50\,Hz. The DRAM reads take \res{dramshare.2334}\% of this energy. The energy of the chip also scales with the bit-steps, from \res{mjchip.w8}\,mJ at uniform 8-bit weights to \res{mjchip.2334}\,mJ at \sch{2,3,3,4}. Most of this reduction comes from the bit-serial sub-arrays, whose cycles scale with the width of each step. \sch{2,3,3,4} needs \res{pct.e2334w8}\% of the energy of uniform 8-bit weights, and \sch{2,4} at two steps needs \res{pct.e24w8}\%. On the accelerator, the per-step precision thus reduces the energy per inference as well as the time.

\begin{figure}[t]
\centering
\includegraphics[width=\columnwidth]{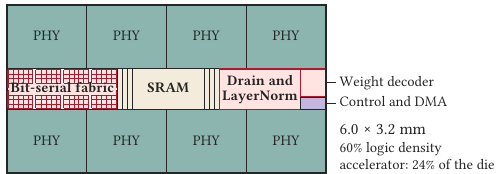}
\caption{A proposed die layout at 7\,nm.}
\label{fig:die}
\end{figure}

Figure~\ref{fig:die} shows a proposed die layout at 7\,nm. We sized the PHYs of the eight channels from the die shot of Apple's M2, a TSMC 5\,nm processor with a memory interface of the same 128-bit width~\cite{m2dieshot}. The PHYs then take \res{mm2.phy}\,\mmsq and the accelerator \res{mm2.placed}\,\mmsq at 60\% logic density, so the PHYs occupy \res{pct.diephy}\% of the \res{die.placed}\,\mmsq die. The accelerator is therefore small compared with the PHYs of its memory interface.

\subsection{Comparison with the State-of-the-Art}
\label{sec:eval-compare}

\begin{figure}[t]
\centering
\includegraphics[width=\columnwidth]{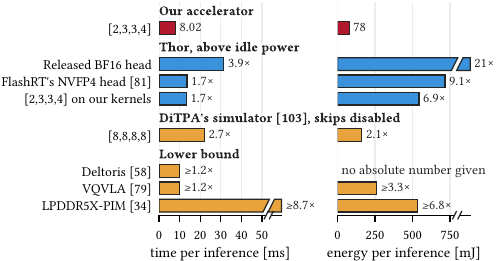}
\caption{State-of-the-art comparison on the GR00T N1.6 action head.}
\label{fig:compare}
\end{figure}

Figure~\ref{fig:compare} compares our accelerator with Thor~\cite{jetsonthor}, recent accelerators for robot policies and LPDDR5X-PIM~\cite{samsungpim}. We scaled the power of DiTPA~\cite{ditpa} to 7\,nm with the factors of Stillmaker and Baas~\cite{stillmaker}. We used our DRAM model for the DRAM energy of every design except Thor, whose energy we measured. Deltoris~\cite{deltoris} and VQVLA~\cite{vqvla} publish no simulator. We therefore bounded their time from below by the transfer time of their DRAM traffic at the peak bandwidth of our memory interface. These bounds favor both designs, since they assume an array of any size and a weight format that preserves the released model's success rate.

\textbf{Thor}~\cite{jetsonthor} is NVIDIA's commercial GPU module for robots, with twice our DRAM bandwidth. Our accelerator nevertheless runs \sch{2,3,3,4} \res{x.thor2334}\xx faster than Thor, and \res{x.flashrt}\xx faster than FlashRT's NVFP4 head~\cite{flashrt} on Thor. Figure~\ref{fig:overview}b shows that Thor is faster only at uniform 8-bit weights, where its DRAM floor lies at half of ours. Scaled to Thor's bandwidth, however, our accelerator is \res{x16ch.w8}\xx faster than Thor even at uniform 8-bit weights, and \res{x16ch.2334}\xx faster at \sch{2,3,3,4}, as Figure~\ref{fig:results}a shows. Our accelerator leads at equal bandwidth because its time stays close to its DRAM floor, whereas fixed overheads keep Thor's time far above its DRAM floor. Figure~\ref{fig:overview}c shows that, above its idle power, Thor uses \res{mj.thor2334}\,mJ per \sch{2,3,3,4} inference, against \res{mj.2334}\,mJ for our accelerator. As shown in Figure~\ref{fig:compare}, FlashRT's NVFP4 head uses \res{xj.flashrt}\xx the energy of our accelerator.

\textbf{DiTPA} (ISCA'26~\cite{ditpa}) is an accelerator for DiT-based action planners that skips redundant actions, flow steps and tokens. With its skips disabled, DiTPA reads 8-bit weights at every flow step and runs \res{x.ditpaw8}\xx faster than our accelerator with uniform 8-bit weights, as Figures~\ref{fig:compare} and~\ref{fig:results}a show. Its INT8 array at 500\,MHz, however, is too slow to keep pace with its DRAM, whose bandwidth is 1.4\xx ours. Narrower weights would therefore not shorten the inference on DiTPA, since its array takes the same number of cycles at every width. DiTPA's skips reduce these cycles but leave the bits that each flow step reads unchanged. Even with all its skips, DiTPA reports \res{hz.ditpan15}\,Hz on GR00T N1.5 and \res{hz.ditpapi25}\,Hz on \pifive. Our accelerator runs GR00T N1.5 at \sch{2,3,3,4} \res{x.ditpan15}\xx faster than DiTPA without its projection of System~2's keys and values. It runs \pifive \res{x.ditpapi25}\xx faster at \sch{2,5}, two flow steps instead of ten. On these action heads, reducing the bit-steps thus shortens the inference more than DiTPA's skips do.

\textbf{Deltoris} (MICRO'26~\cite{deltoris}) is an accelerator for diffusion VLAs that computes each linear layer from the nonzero bits of its input change since the previous call. Its speculation on predicted frames does not apply to GR00T N1.6, which predicts no frames. To compute the input changes, Deltoris keeps the inputs and outputs of every linear layer of the previous call in DRAM. On GR00T N1.6, this state far exceeds the on-chip SRAM, so reading and writing it adds DRAM traffic to the weight reads. For the weights of Deltoris, our bound assumes the 2-bit vector code of Section~\ref{sec:eval-quant}, which reads as many bits as coded \sch{2,3,3,4}. Even with an array of any size, Deltoris therefore takes longer than our accelerator. Computing on input changes thus saves arithmetic but adds DRAM traffic, which lengthens the inference of a DRAM-bound action head.

\textbf{VQVLA} (MICRO'26~\cite{vqvla}) is an accelerator for VLAs that reads one of two vector-quantized copies of the weights, chosen per inference from the last arm motion. Even its narrower copy reads 12.5 bit-steps, against \res{sota.ours.bits} for coded \sch{2,3,3,4}. On a DRAM-bound action head, VQVLA thus remains slower than our accelerator even with an array of any size. Its index units and result cache, which merge and reuse the multiplications by shared centroids, take 86\% of its 51.2\,\mmsq and 88\% of its power. This reuse saves arithmetic but no DRAM traffic, on which the time of a DRAM-bound action head depends.

\textbf{LPDDR5X-PIM}~\cite{samsungpim} is Samsung's fabricated LPDDR5X DRAM with MAC units at its banks, which perform only one MAC per weight read (Section~\ref{sec:ondram}). At their peak rate, the \res{gmac.2334} GMAC of a \sch{2,3,3,4} inference take at least \res{x.pim}\xx the time of our accelerator. Because LPDDR5X-PIM reads each weight in its bank once per MAC, these reads also take more energy than our whole inference.

Of the designs compared, only our accelerator thus combines the fewest bit-steps with a time close to their transfer time.

\subsection{Programmability Across Models}
\label{sec:eval-vlm}

\begin{figure}[t]
\centering
\includegraphics[width=\columnwidth]{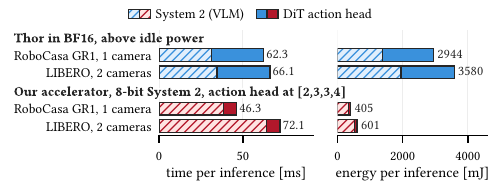}
\caption{Time and energy of the whole policy.}
\label{fig:policy}
\end{figure}

\textbf{Other action heads.} As Figure~\ref{fig:results}a shows, the same RTL runs \sch{2,3,3,4} on GR00T N1.5 in \res{ms.n15}\,ms per inference and \sch{2,5} on \pifive at two flow steps in \res{ms.pi25}\,ms. One accelerator thus runs action heads of different sizes and architectures, each above 50\,Hz.

\textbf{Vision-language model.} Our accelerator's fabric is flexible enough to also run System~2, the vision-language model of the policy. We ran the System~2 of GR00T N1.6 on the RTL with 8-bit weights and INT8 inputs, bit-exact against the executable model of the accelerator. On the accelerator, a policy inference runs System~2 followed by the action head at \sch{2,3,3,4}. Figure~\ref{fig:policy} compares these inferences with NVIDIA's VLM backbone and the released BF16 head on Thor.

On a RoboCasa GR1 call with one camera, the accelerator runs the action head \res{x.headthor}\xx faster than the released BF16 head on Thor. System~2, however, takes \res{x.s2thor}\xx as long as on Thor. System~2 is compute-bound, so Thor is faster because it has much higher peak compute. On the one-camera call, a policy inference costs \res{s2.rcpolicymj}\,mJ on the accelerator, against \res{thor.policymj}\,mJ above idle power on Thor.

At an action-head rate of 50\,Hz, \sch{2,3,3,4} on the one-camera call leaves \res{s2.free}\,ms of every 20\,ms period free. Since System~2 can pause at layer boundaries, its inference fits into the free time of \res{s2.periods} action-head inferences. System~2 can thus run at \res{s2.hz}\,Hz, faster than the 10\,Hz or less of dual-system policies~\cite{gr00tn1,helix,leverb,hume}.

\insight{By speeding up the action head, the per-step precision frees a programmable accelerator to also run System~2 between action-head inferences.}

\subsection{Quantization-in-DRAM}
\label{sec:eval-ondram}

\begin{figure}[t]
\centering
\includegraphics[width=\columnwidth]{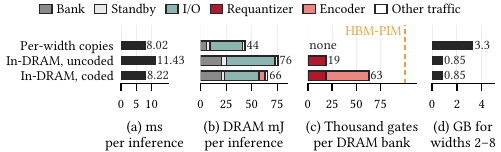}
\caption{Quantization-in-DRAM against per-width copies.}
\label{fig:ondram}
\end{figure}

On one \sch{2,3,3,4} inference of GR00T N1.6, Figure~\ref{fig:ondram} compares two options of Quantization-in-DRAM with the per-width copies that all other experiments read. With requantizers alone, the DRAM transfers every width uncoded. With an encoder behind every requantizer, it sends the weights in the Huffman code of the per-width copies. We verified that the RTL of the requantizer and the encoder is bit-exact with these copies and synthesized it in TSMC 7\,nm at the 3.75\,ns column cycle of LPDDR5X-8533. We scaled both the area of the netlists, placed at 60\% density, and their energy by the 10\xx lower density of logic in a DRAM process (Section~\ref{sec:ondram}).

Only with the encoders does Quantization-in-DRAM keep the time per inference close to that of the per-width copies. Without them, every width crosses the interface uncoded, so an inference takes \res{x.odruncoded}\xx as long, as Figure~\ref{fig:ondram}a shows. With them, an inference takes \res{pct.odrrtl}\% longer, since the banks' tags and padding add \res{pct.odrtraffic}\% to the DRAM traffic. In a DRAM timing model that we validated against Ramulator~2, the reads of the 8-bit copy inside the banks add another \res{pct.odrtime}\%. With the encoders, Quantization-in-DRAM raises the DRAM energy by \res{pct.odrbank}\% (Figure~\ref{fig:ondram}b), mainly because every bank reads all 8 bits of every weight instead of only the coded bits.

Samsung's HBM-PIM~\cite{hbmpim}, a fabricated chip, places logic at each bank of its 20\,nm DRAM. Figure~\ref{fig:ondram}c shows that the requantizer and encoder of a bank need fewer gates than this logic. Their critical paths in 7\,nm take at most \res{ns.odrpath}\,ns of the column cycle, so they meet it even with three times slower DRAM transistors~\cite{upmem}. The requantizers and encoders take \res{pct.odrdie}\% of a two-channel 16\,Gb LPDDR5X die of 37\,\mmsq~\cite{fmsdram}, or \res{pct.odrdiepim}\% at the area per gate of HBM-PIM. A DRAM die can therefore hold a requantizer and an encoder at every bank.

With encoders at the banks, Quantization-in-DRAM thus trades DRAM energy and die area for capacity, at almost the time per inference of the per-width copies.

\section{Related Work}
\label{sec:related}

\textbf{Quantized robot policies.} Prior work quantizes current VLAs to uniform 4-bit weights, with plain rounding and with calibrated methods~\cite{loopq,foldquantvla,chasevla,vlaquantbench,holoqvla}. SACQ~\cite{sacq} preserves success on LIBERO at 2 bits per weight with vector codebooks trained for every suite. BitVLA~\cite{bitvla} trains a ternary backbone and ActQuant~\cite{actquant} rounds the backbone to about 3 bits from calibration demonstrations, both with a full-precision action head. QbD-VLA~\cite{qbdvla} selects the widths of layer groups by short closed-loop rollouts. QuantWAMs~\cite{quantwams} and DyQ-VLA~\cite{dyqvla} vary the activation precision per step and per policy query, respectively, and SteerQuant~\cite{steerquant} keeps static 4-bit weights. None of these works varies the weight width over the flow steps of one inference. The methods that hold below 4 bits train or calibrate on target data, whereas our schedules need neither.

\textbf{Time-aware quantization of diffusion models.} Quantizers for image diffusion calibrate activation ranges per timestep~\cite{qdiffusion,ptq4dm,ptq4dit,tdq}, mix precision per layer~\cite{mixdq} or choose the steps together with a per-layer precision~\cite{tmpqdm,diffpro}. PTQD~\cite{ptqd} raises the activation width over the steps, and AdaTSQ~\cite{adatsq} and MASQ~\cite{masq} vary it per step or per image region. Our rising widths are those of the weights, which determine the DRAM traffic at batch size one. TMPDiff~\cite{tmpdiff} and TASQ~\cite{tasq} vary the weight precision per step on images (Section~\ref{sec:gains}). The analysis of TASQ sets the Jacobian of the denoiser to zero and thereby ignores how the remaining steps attenuate an early error. Let It Be Simple~\cite{letitbesimple} shows that an action head's flow attenuates differences in its starting noise, whereas the per-step gains and the rising widths derived from them are new.

\textbf{Step counts.} One or a few flow steps often suffice for action heads~\cite{letitbesimple,onestep,kmf}, depending on how many actions run per inference~\cite{onestep,coupling}. Several works choose the step count per inference~\cite{adavla,vlarolling} or distill the head to one or two steps~\cite{snapflow,realtimevla}. We treat the step count, like the widths, as a parameter.

\textbf{Accelerators for robot policies.} SpecVLA~\cite{specvla} verifies long chunks with the same network applied to quantized changes of its inputs. DaDu-Corki~\cite{corki} predicts a trajectory per inference, so that the VLA runs less often. Both run every full inference of the policy on a GPU. SpecVLA's hardware runs the verifier, and DaDu-Corki's converts the predicted trajectories into joint torques. They reduce the number of inferences, which is orthogonal to reducing the time and energy of each.

\textbf{Bit-serial and precision-scalable arrays.} Bit-serial and bit-flexible designs scale their cycles with the operand widths~\cite{stripes,pragmatic,loom,bitfusion,bismo,unpu,bitwave}. BitMoD~\cite{bitmod} serializes the weights of language models at 3 to 8 bits, and low-bit formats adapt the data type to the values~\cite{ant,olive}. Run-time partitioning divides an array between tenants by their deadlines~\cite{planaria}. To the best of our knowledge, no design partitions an array at run time by what a memory-bound stream needs.

\textbf{Logic in memory.} Processing-in-memory DRAMs compute beside the banks with one MAC per weight read~\cite{hbmpim,aim,samsungpim,cdpim} or evaluate lookup tables inside the sub-arrays~\cite{pluto}. An extension of HBM-PIM~\cite{kimpim} and StreamDQ~\cite{streamdq} widen quantized weights to 16 bits, for the MAC units or in the logic base die of an HBM stack. JIT-Q~\cite{jitq} quantizes a master copy with the PIM units of HBM and writes the result back for the GPU. In a CXL memory device, SmartQuant~\cite{smartquant} and TRACE~\cite{trace} read only the top bits of one stored copy, which TRACE compresses with a generic codec. The closest DRAM system is WIFiV-LPDDR~\cite{bdllm}, a modeled LPDDR that narrows a stored 8-bit copy on read by a clamp and a shift. It neither specifies where in the DRAM this narrowing takes place nor encodes the weights, whereas we place a requantizer and an encoder at every bank.

\section{Conclusion and Future Work}
\label{sec:conclusion}

Every flow step of a robot action head re-reads all of its weights but applies each to only about fifty tokens, so its inference is memory-bound. Because later flow steps attenuate early quantization errors, the head's weight precision should rise over the steps. With plain per-channel rounding, \sch{2,3,3,4} preserves the closed-loop success of three policies, but current GPU kernels cannot exploit this per-step precision due to fixed overheads. We therefore propose a programmable accelerator whose time and energy scale with the bit-steps, because it keeps its DRAM link saturated throughout an inference at every width. It decodes entropy-coded weights at the DRAM rate for eight bit-serial sub-arrays, which also run the VLM. A run-time allocator divides them between the linear layers and the attention, whose compute requirements change with the width and the model. Quantization-in-DRAM, with a requantizer and an encoder at each bank, lets one stored 8-bit copy serve every width. With a copy per width and 50\% of Thor's memory bandwidth, the accelerator runs GR00T N1.6 at \sch{2,3,3,4}. An inference then completes \res{x.flashrt}\xx faster than on Thor in NVFP4, at \res{xj.flashrt}\xx less energy and equal closed-loop success. On a one-camera call, one inference per action at 50\,Hz takes \res{pct.period2334}\% of each control period and leaves time to run the VLM at \res{s2.hz}\,Hz.

\textbf{Future work.} A run-time selector such as that of VQVLA~\cite{vqvla} could set the widths of every inference as the battery, the thermal budget or the task demands.

\textbf{Limitations.} Every closed-loop result comes from simulation, and quantized policies that hold in simulation have lost success on real robots~\cite{vlaquantbench,qbdvla}. Quantization-in-DRAM needs a DRAM mode that no commercial part offers. The energy of the chip comes from VCD-annotated netlists with a clock tree but before placement and routing, and that of the DRAM from a model.

\begin{acks}
This project has been partly funded by the European Research Council (ERC) under grant agreement No.~101088865, the Flanders AI Research Program, Research Foundation Flanders (FWO) under grant No.~1S37125N, and long-term structural Methusalem funding by the Flemish Government. We acknowledge NVIDIA Corporation for awarding GPU hardware through its Academic Grant Program.
\end{acks}

\bibliographystyle{ACM-Reference-Format}
\bibliography{references}

\end{document}